\documentclass[12pt]{iopart}
\usepackage{harvard}
\usepackage{bm}
\usepackage{graphicx}
\graphicspath{{fig/}}

\newcommand{\deriv}[3]{\frac{#3\hspace*{-.06em} {#1}}{#3\hspace*{.06em} {#2}}}

\newcommand{\parder}[2]{\deriv{#1}{#2}{\partial}}

\newcommand{\EQ}{\begin{equation}}
\newcommand{\EN}{\end{equation}}
\newcommand{\EQA}{\begin{eqnarray}}
\newcommand{\ENA}{\end{eqnarray}}
\newcommand{\eq}[1]{(\ref{#1})}

\newcommand{\Eq}[1]{equation~(\ref{#1})}

\newcommand{\Sec}[1]{\S\ref{#1}}

\newcommand{\Fig}[1]{Fig.~\ref{#1}}
\newcommand{\FFig}[1]{Figure~\ref{#1}}

\newcommand{\bra}[1]{\langle #1\rangle}

\newcommand{\mean}[1]{\overline #1}
\newcommand{\meanrho}{\overline{\rho}}

\newcommand{\meanemf}{\overline{\mbox{\boldmath ${\cal E}$}}{}}{}
{}
\newcommand{\meanemfs}{\overline{\cal E} {}}

{}
\newcommand{\aTens}{\mbox{\boldmath $\alpha$}}

\newcommand{\eTens}{\mbox{\boldmath $\eta$}}
{}
\newcommand{\meanEE}{\overline{\mbox{\boldmath ${\cal E}$}}{}}{}
\newcommand{\meanEMF}{\overline{\mbox{\boldmath ${\cal E}$}}{}}{}
{}
{}
{}
{}
{}
\newcommand{\meanBB}{\overline{\mbox{\boldmath $B$}}{}}{}
\newcommand{\meanJJ}{\overline{\mbox{\boldmath $J$}}{}}{}
\newcommand{\meanUU}{\overline{\mbox{\boldmath $U$}}{}}{}
{}
{}

\newcommand{\meanB}{\overline{B}}

{}
\newcommand{\hatBB}{\hat{\mbox{\boldmath $B$}}{}}{}
{}
{}

\newcommand{\hatB}{\hat{B}}
\newcommand{\alphaK}{\alpha_{\rm K}}
\newcommand{\alphaM}{\alpha_{\rm M}}
\def\hf{h_{\rm f}}
\def\hm{h_{\rm m}}
\newcommand{\kk}{\mbox{\boldmath $k$} {}}

\newcommand{\uu}{\mbox{\boldmath $u$} {}}
\newcommand{\UU}{\mbox{\boldmath $U$} {}}
\newcommand{\xx}{\mbox{\boldmath $x$} {}}
\newcommand{\aaa}{\mbox{\boldmath $a$} {}}
\newcommand{\bb}{\mbox{\boldmath $b$} {}}
\newcommand{\cc}{\mbox{\boldmath $c$} {}}
\newcommand{\BB}{\mbox{\boldmath $B$} {}}

\newcommand{\jj}{\mbox{\boldmath $j$} {}}
\newcommand{\JJ}{\mbox{\boldmath $J$} {}}

\newcommand{\AAA}{\mbox{\boldmath $A$} {}}

\newcommand{\nab}{\mbox{\boldmath $\nabla$} {}}
\newcommand{\OO}{\mbox{\boldmath $\Omega$} {}}

\newcommand{\ii}{{\rm i}}

\newcommand{\curl}{{\rm curl} \, {}}

\newcommand{\dd}{{\rm d} {}}

\def\St{\mbox{\rm St}}

\def\Pm{\mbox{\rm Pr}_M}
\def\Rm{\mbox{\rm Re}_M}

\def\kf{k_{\rm f}}

\def\etat{\eta_{\rm t}}

\def\Beq{B_{\rm eq}}
\newcommand{\ea}{{\it et al\/ }}

\newcommand{\yapj}[3]{ #1, {\it Astrophys.\ J.} {\bf #2} #3}
\newcommand{\yapjs}[3]{ #1, {\it Astrophys.\ J.\ Suppl.} {\bf #2} #3}
\newcommand{\yan}[3]{ #1, {\it Astron.\ Nachr.} {\bf #2} #3}

\newcommand{\yana}[3]{ #1, {\it Astron. Astrophys.} {\bf #2} #3}
\newcommand{\ygafd}[3]{ #1, {\it Geophys.\ Astrophys.\ Fluid Dyn.} {\bf #2} #3}
\newcommand{\yjfm}[3]{ #1, {\it J.\ Fluid Mech.} {\bf #2} #3}

\newcommand{\ysov}[3]{ #1, {\it Sov.\ Astron.} {\bf #2} #3}

\newcommand{\yprs}[3]{ #1, {\it Proc.\ Roy.\ Soc.\ Lond.} {\bf #2} #3}

\newcommand{\yprl}[3]{ #1, {\it Phys.\ Rev.\ Lett.} {\bf #2} #3}

\newcommand{\yptrs}[3]{ #1, {\it Phil.\ Trans.\ Roy.\ Soc.} {\bf #2} #3}
\newcommand{\ymn}[3]{ #1, {\it Mon.\ Not.\ R.\ Astron.\ Soc.} {\bf #2} #3}
\newcommand{\ynat}[3]{ #1, {\it Nature} {\bf #2} #3}

\newcommand{\yjour}[4]{ #1, {\it #2}, {#3} #4}

\newcommand{\ybook}[3]{ #1, {#2} (#3)}
\newcommand{\yproc}[5]{ #1, in {#3}, ed.\ #4 (#5) #2}

\newcommand{\sana}[1]{ #1, {\it Astron. Astrophys.} (submitted)}

\usepackage{iopams}  

\begin{document}

\title[Turbulent transport in hydromagnetic flows]{Turbulent transport in hydromagnetic flows}

\author{A Brandenburg$^{1,2}$, P Chatterjee$^1$, F Del Sordo$^{1,2}$,
A Hubbard$^1$, P J K\"apyl\"a$^{1,3}$, M Rheinhardt$^{1}$}

\address{1. NORDITA, Roslagstullsbacken 23, SE-10691 Stockholm, Sweden}
\address{2. Department of Astronomy, Stockholm University, SE-10691 Stockholm, Sweden}
\address{3. Department of Physics, FI-00014 University of Helsinki, Finland}
\ead{brandenb@nordita.org, $ $Revision: 1.174 $ $}

\begin{abstract}
The predictive power of mean-field theory is emphasized by comparing
theory with simulations under controlled conditions.
The recently developed test-field method is used to extract
turbulent transport coefficients both in kinematic as well as
nonlinear and quasi-kinematic cases.
A striking example of the quasi-kinematic method is provided by
magnetic buoyancy-driven flows that produce an $\alpha$ effect
and turbulent diffusion.
\end{abstract}


\section{Introduction} 
What happens when fluids mix?
What if a fluid is moving in a magnetized environment?
Are there analogies between the motion of a cloud in the sky, milk in
a coffee cup and the solar flares?
The study of fluids and magnetic fields has always been a challenging branch
of physics, leading to the development of tools of wide applicability,
from meteorology to the study of galaxies.
In particular the connection between the existence of fluids in motion
and the amplification of magnetic fields has been investigated
both analytically and experimentally since the
beginning of the twentieth century.
The generation of a magnetic field by dynamo action was already proposed
by Larmor (1919), but a proper understanding of such a process
requires both physical insight and a theoretical framework that describes
the magneto-hydrodynamical (MHD) context in which the phenomena occur.
The most common theoretical approach to MHD dynamos is the application of
mean-field theory (Parker 1955, Steenbeck and Krause 1969, Moffatt 1978,
Parker 1979, Krause and R\"adler 1980).
The core concept on which mean-field theory (hereafter MFT)
rests is that turbulent systems (which include
most natural MHD dynamos) are often amenable to a two-scale approach,
where the velocity and magnetic fields
are decomposed into mean
and fluctuating components: $\UU = \meanUU + \uu$ and $ \BB =\meanBB + \bb$.
The mean parts $\meanUU$ and $\meanBB$ generally vary slowly
both in space and time, and capture the global,
and often observable
behavior of the system.
The fluctuating fields on the other hand describe
irregular, often chaotic small-scale effects.

Using the aforementioned decomposition
the equation for the time evolution of the magnetic field,
known as the induction equation, can be rewritten as a set
of two equations for mean and fluctuating quantities,
\begin{eqnarray}
&&{\partial\meanBB\over\partial t}=\nab\times\left(\meanUU\times\meanBB\right)
+\nabla\times\meanemf +\eta\nabla^2\meanBB,
\label{dBdt} \\
&&{\partial\bb\over\partial t}=\nab\times\left(\meanUU\times\bb\right)
+\nab\times\left(\uu\times\meanBB\right)+\nab\times\left(\uu\times\bb\right)'
+\eta\nabla^2\bb,
\label{dbdt}
\end{eqnarray}
where $\eta$ is the microphysical magnetic diffusivity of the fluid
(here assumed uniform), while $\meanemf\equiv\overline{\uu\times\bb}$
is the mean electromotive force, and
$\left(\uu\times\bb\right)'=\uu\times\bb -\overline{\uu\times\bb}$.

Correlations such as $\overline{\uu\times\bb}$ are at the heart of turbulent
transport, and apply to a broad range of processes, from dynamos to the
mixing of chemicals through stirring.
Here, the key task consists in relating $\meanemf$ to the mean field $\meanBB$.
Underlying symmetries that constrain the form of this relation
are a significant help.
$\meanemf$ is a vector
so if the system is homogeneous and the turbulence isotropic,
in what direction can it point?
The answer is that in such a system $\meanemf$ can have constituents
pointing along
the mean magnetic field $\overline{\BB}$ and the mean current density
$\overline{\JJ}=\nab\times\meanBB/\mu_0$
(as well as higher order spatial and
time derivatives, see \Sec{Nonlocality}), leading
to approximations such as
\EQ
\meanemf=\alpha \overline{\BB}-\etat \mu_0\overline{\JJ}.
\label{emf}
\EN
The coefficients linking correlations to mean quantities are known 
unimaginatively as mean-field transport coefficients, with each one
describing a distinct physical effect.
In \Eq{emf}, $\alpha$ describes the (in)famous $\alpha$ effect which
can drive a dynamo while $\etat$ quantifies the turbulent diffusion of
the mean magnetic field, and $\mu_0$ is the vacuum permeability.
Note that a much more general representation of $\meanemf$ is given 
by the convolution integral
\EQ
   \meanemf(\xx,t) = \int_{t_0}^t \int \boldsymbol{G}(\xx,\xx',t,t')\, \meanBB(\xx',t') \, \dd^3 x' \, dt' \label{convol}
\EN
with an appropriate tensorial kernel $\boldsymbol{G}$.

Equation~(\ref{dbdt}) contains
terms that can sometimes be neglected.
Most famously, in the case of fluids
with small magnetic Reynolds number, that is $\Rm=UL/\eta\ll1$,
or low Strouhal number $\St=U\tau_{\rm c}/L\ll 1$ we can drop
$(\uu\times\bb)'$ in \Eq{dbdt} and can thus make an analytical
calculation of the transport coefficients feasible.
Under this approximation, known as SOCA (Second Order
Correlation Approximation), \Eq{dbdt} takes the form
\EQ
{\partial\bb\over\partial t}=\nab\times\left(\meanUU\times\bb\right)
+ \nab\times\left(\uu\times\meanBB\right) + \eta\nabla^2\bb.
\label{dbdtsoca}
\EN
In the limit of high $\Rm$ (hence small $\St$)
the coefficients $\alpha$ and $\eta_t$
reduce then to (Krause and R\"adler 1980, R\"adler and Rheinhardt 2007)
\EQ
\alpha=-\frac{\tau_{\rm c}}{3}\,\overline{\uu\cdot\left(\nab\times\uu\right)},
\quad \etat=\frac{\tau_{\rm c}}{3}\overline{\uu^2},
\label{alpha}
\EN
where $\tau_{\rm c}$ is a characteristic turbulent correlation time.

\section{The need for MFT: a status report}

\subsection{Motivation}

In the astrophysical context,
MFT has mainly been applied in
order to understand and model the origin of the Sun's magnetic field
and its differential rotation (R\"udiger and Hollerbach 2004).
Direct simulations of convection in spherical shells begin
to reproduce these phenomena to some extent
(Brun \ea 2004, Browning \ea 2006, Brown \ea 2010, K\"apyl\"a \ea 2010a),
but interpreting their results properly remains difficult.
This task is approachable only in the framework
of a reasonably accurate theory.

MFT is sometimes perceived as uncertain and even arbitrary
owing to a large amount of parameters that are often chosen to reproduce
``whatever one wants''.
Adjusting parameters at will is certainly risky and
clearly not permissible in the long run, because it
would imply a complete loss of predictive power of MFT.
There are several reasons why the ``free parameter'' approach has often been adopted.
Firstly, the conventional theory for computing turbulent transport coefficients
is only accurate at low Reynolds numbers and not well tested at
larger Reynolds numbers.
Secondly, models of solar-like dynamos that are based on a
straight-forward application of mixing-length ideas to computing
turbulent transport coefficients (Krivodubskii 1984) do not reproduce the Sun:
the cycle periods are too short (K\"ohler 1973) and the migration of sunspot
and other magnetic activity is poleward, not equatorward, which is also
found in direct numerical simulations (Gilman 1983, K\"apyl\"a \ea 2010a).

In this situation it is sensible to reduce ones ambitions 
and to focus on phenomena that are seen in direct simulations of
simplified systems which are nevertheless relevant for understanding the Sun.
A useful goal consists then in reproducing such phenomena
by mean-field models, thus obtaining a chance to trace
down the reasons for discrepancies between both representations.
This will be exemplified in \Sec{Predictions}. 
First, however, we shall summarize the basic saturation phenomenology
of mean-field dynamos.

\subsection{Saturated dynamos and magnetic helicity fluxes}
\label{helicity}
A recent discovery that is now well explained by MFT is the slow
saturation behavior of an $\alpha^2$ dynamo in a triply periodic box
(Brandenburg 2001).
Such systems are
unphysical, but they make good test problems due to the ease of
capturing them both numerically and analytically.
For early times, both the
mean and the fluctuating fields grow exponentially -- as expected
from kinematic theory.
However, when the small-scale
field becomes comparable to the equipartition value,
i.e.\ $\overline{\bb^2}\sim\Beq^2$,
the growth changes its nature: the fluctuating field saturates while the mean
field, well below equipartition, continues to grow albeit extremely slowly.
Finally, after multiple microphysical resistive times, the mean field
itself reaches a steady, super-equipartition state;
see Fig.~9.4 of Brandenburg and Subramanian 2005).

This behavior is one aspect of the ``catastrophic'' $\alpha$-quenching
phenomenon, and has come to be understood in terms of the magnetic
helicity density, $h\equiv \AAA \cdot \BB$, and the magnetic $\alpha$
effect of Pouquet \ea (1976), where the growing magnetic $\alpha$,
$\alphaM \equiv \tau \overline{\jj\cdot\bb}/3\meanrho$, is subtracted
from the $\alpha$ of \Eq{alpha}, now marked $\alphaK$ (kinetic), so
that the net $\alpha=\alphaK+\alphaM$ would be reduced and the dynamo growth
halted.  This result has been extended
to the dynamic $\alpha$ quenching phenomenology
(Kleeorin and Ruzmaikin 1982, Field and Blackman 2002,
Blackman and Brandenburg 2002), where the mean magnetic helicity
in the fluctuating fields, $\hf \equiv \overline{\aaa \cdot \bb}$,
is used as a proxy for the
current helicity, $\overline{\jj \cdot \bb } \simeq \kf^2 \hf$,
with $\kf$ being the wavenumber of the energy-carrying eddies.
In a system that disallows the divergence of the magnetic helicity flux,
such as a triply periodic domain, the time evolution of $\hf$ and the
resulting dynamical $\alpha$ quenching equation can be written as:
\begin{eqnarray}
&&\frac{{\rm d} \hf}{{\rm d} t}=-2\meanemf \cdot \overline{\BB}
-2\eta \overline{\jj \cdot \bb},
\\
&&\frac{\rm{d} \alpha}{\rm{d} t}=-2\etat \kf^2 \left(\frac{\alpha \overline{\BB}^2
-\eta_t \overline{\JJ} \cdot \overline{\BB}}{\Beq^2}+\frac{\alpha-\alphaK}{\Rm}\right).
\end{eqnarray}
The three phases of the $\alpha^2$ dynamo in a triply periodic domain
can now be understood.
First the fields grow exponentially and the magnetic $\alpha$ effect
grows with them until $\alphaM \sim -\alphaK$.
This occurs rapidly enough that still $\hm \sim -\hf$ and so
$\meanB/\Beq\simeq\!\sqrt{k_1/\kf}<1$ (Brandenburg 2001),
where $k_1$ is the smallest possible wavenumber in the domain.
During the resistive phase, the fluctuating fields are nearly
steady but there is still a small excess of $\alpha$ over $\eta k_1$
needed to replenish the field
in the face of resistivity. 
This phase ends only when the time evolution of
the total magnetic helicity reaches a steady state, which occurs when
$\langle \JJ \cdot \BB\rangle=0$, or for
$\meanB/\Beq\simeq\sqrt{\kf/k_1}>1$.

A short exponential growth phase yielding only
weak mean fields poses severe problems
for astrophysics as the subsequent
resistive growth phase is generally prohibitively long.
Real systems however allow for fluxes of magnetic helicity
across their borders, and/or show 
spatial variations in the $\alpha$ effect, particularly
regions where $\alpha$ has opposite signs
separated, say, by an equator.
This raises the possibility that the magnetic $\alpha$ will be exported
from the system or transported to the equator and destroyed.
Research into such transport is recent, but has already shown conclusively
that there is a flux of $\hf$ and that a larger residual $\alpha$ effect
results due to it.
It is not yet clear how large an effect this has on the
final mean field strength.

\section{Predictions versus realizations}
\label{Predictions}

In this section we contrast the results of some computer realizations
with corresponding mean-field predictions.
We discuss examples from both linear and nonlinear regimes.

\subsection{Parity and dependence on boundary conditions}

A relatively old example is the emergence of oscillatory dynamo
solutions in local models of accretion discs (Brandenburg \ea 1995).
Here, turbulence is driven by the magneto-rotational instability which
generates a negative $\alpha$ effect in the upper half of the disc
(Brandenburg \ea 1995, Ziegler and R\"udiger 2000,
Brandenburg 2005a, Gressel 2010).
According to MFT, this negative $\alpha$, together with
pseudo-vacuum boundary conditions (that is, normal field conditions),
predicts traveling wave solutions that are symmetric about the midplane
and migrating toward the boundaries (Brandenburg and Campbell 1997).
Conversely, when the boundary condition is changed to a perfect conductor
condition, one expects non-oscillatory solutions that are
antisymmetric about the midplane.
Again, this dependence is born out by simulations (Brandenburg 1998).

\subsection{Onset of convective dynamo action depending on rotation rate}

Large-scale dynamos due to turbulent convection are of particular
interest in astrophysics. According to MFT, rotating
inhomogeneous (usually due to stratification) turbulence leads to the
generation of kinetic helicity and thus an $\alpha$ effect which should
enable the generation of large-scale fields.
However, numerical simulations of rotating convection
at first failed to show large-scale dynamo
action (e.g.\ Brandenburg \ea 1996, Cattaneo and Hughes 2006).
Erroneously low values of $\alpha$ determined with what is now often called
the imposed-field method, where a uniform magnetic field $\BB_0$ is applied
and one determines, from the simulation, the mean electromotive force,
$\bra{\uu\times\bb}$, where mean fields are defined as volume averages.
(e.g.\ Cattaneo and Hughes 2006; see, however, K\"apyl\"a \ea
2010b) seemed to confirm that the $\alpha$ effect does not work.
On the other hand, when computing turbulent transport coefficients for
convection using the test-field method (see \Sec{TestField})
it was discovered that, as the rotation
rate in non-shearing runs increases, the $\alpha$ effect increases and
turbulent diffusion, $\etat$, decreases (K\"apyl\"a \ea 2009a).
Mean-field models (hereafter MFM), using
these properly determined transport coefficients then
suggested that a large-scale dynamo should be excited when
the Coriolis number, defined as the ratio of the rotation period and
the convective turnover time exceeds a value of $\approx 4$.
Subsequently, direct simulations
in this parameter range just confirmed the existence of a large-scale dynamo
(K\"apyl\"a \ea 2009b).
Again, this demonstrates that already kinematic
MFT has predictive power and that very likely MFMs
can give useful and new information even about more complex systems.

\begin{figure}
\includegraphics[width=\columnwidth]{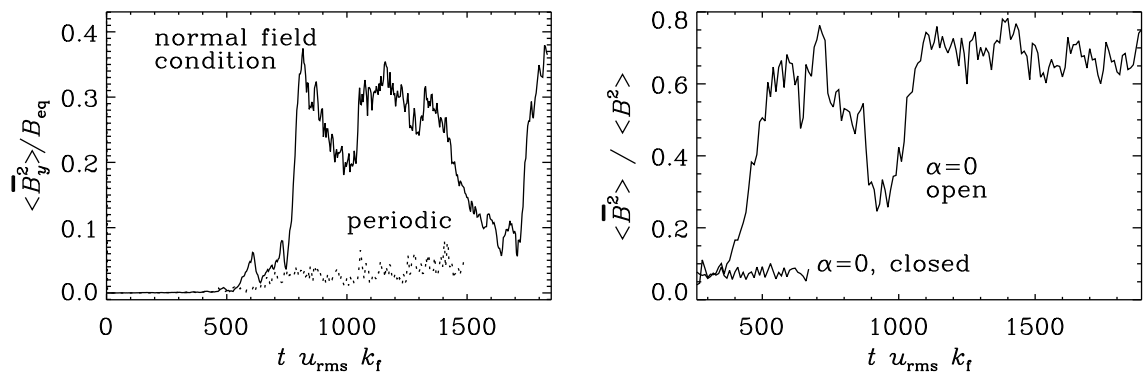}
\caption{
Left: Energy in the horizontally averaged streamwise magnetic field 
from two convection simulations with vertical shear $\mean{U}_y(z)$
and either normal field (solid line) or periodic (dotted line)
magnetic boundary conditions in the $x$ direction.
Adapted from K\"apyl\"a \ea (2008).
Right: Evolution of the ratio of the energy contained in the large-scale field
to the total magnetic energy for open
(vertical field) and closed (perfect conductor) boundaries.
Note that large-scale dynamo action is only possible with open boundaries.
Adapted from Brandenburg (2005b).
}
\label{helflux}
\end{figure}

\subsection{Helicity considerations}

The helicity considerations outlined in \Sec{helicity} provide further
predictive power. The shear-induced (non-diffusive) magnetic helicity flux,
introduced by Vishniac and Cho (2001), has been particularly important
in explaining the existence or the absence of a large-scale
dynamo. For example, Tobias \ea (2008) presented simulations of
convection with vertical shear where no large-scale dynamo was excited
although the necessary ingredients (inhomogeneity, rotation and shear)
were all present. However, in that case the shear-driven
magnetic helicity flux is directed along
the periodic $x$ direction and no net flux out of the system could occur.
Using instead, in an otherwise similar setup, normal field boundary conditions,
which do allow a net flux, K\"apyl\"a \ea (2008) showed that a
large-scale dynamo does exist and indeed saturates
at near-equipartition field strengths; see \Fig{helflux}, where we
also show the effect of open versus closed boundaries
for forced turbulence (Brandenburg 2005b).

Another issue approachable through magnetic helicity considerations
is the convergence problem of the MRI,
i.e.\ the steep decrease of the turbulence level
at low magnetic Prandtl numbers,
$\Pm$, in fully periodic setups (e.g.\ Fromang \ea 2007).
An otherwise similar setup, however, that does allow
a magnetic helicity flux through the vertical boundaries produces
indeed strong large-scale dynamo action, roughly independent
of the value of $\Pm$ (K\"apyl\"a and Korpi 2010).

\section{Computing mean-field transport coefficients}

In view of such success stories there should be an unbroken interest in MFMs
both because of their descriptive capabilities
and their predictive potential,
but we have to realize that there are serious shortcomings of MFT
that have persuaded many researchers to (re)turn to global
direct numerical simulations instead of designing improved MFMs.
This critical stage of MFT can be characterized by the following observations:
\begin{itemize}
\item The limitations of analytic approaches to calculating mean-field coefficients are clearly too restrictive as the interest has moved from pointing out
the qualitative existence of certain effects
to the quantitative reproduction and prediction of such processes. This is due to 
\begin{itemize}
\item the obvious insufficiency of SOCA in astrophysical contexts as usually $\Rm\gg1$ and $\St \not\ll1$
\item the unclear aspects of closure approaches like the $\tau$ approximation (R\"adler and Rheinhardt 2007) 
\item the need of knowledge of velocity correlators like $\overline{u_i u_j \ldots u_n}$ even in mathematically well established (systematic) higher-order correlation approximations
\end{itemize}
\item It is obvious that MFMs for realistic setups with predictive
abilities need to employ transport coefficients that are
\begin{itemize}
\item fully tensorial
\item position dependent
\item dependent on the mean quantities themselves, i.e.\ nonlinear
\item non-local and non-instantaneous
\item including magnetic background fluctuations
\end{itemize}
There is no longer any chance for obtaining powerful models by employing
a few scalar coefficients, the basic structure of which can be derived
analytically leaving a few free parameters to be adapted properly.
\end{itemize}
To find a way out of this impass\'e one might look at how in other fields
of physics/engineering, modeling and simulation of rather complex systems
are being made possible
if the full resolution of the microphysics is not affordable. Let us choose as an example the mechanics of elastic materials, say metals.
Their elastic properties can in principle be derived from the microphysics of their lattices, but it will perhaps never be possible to simulate the
static and dynamic behavior of, say, a bridge by solving quantum physical lattice equations. Instead, one relies upon the equations of continuum mechanics
in which the lattice physics enters via macrophysical material properties like Young's modulus and Poisson number (sufficient for an isotropic material). 
These are typically obtained by measurements in a series of standardized experiments with test bodies having simple geometries and being subject to 
clearly defined boundary conditions. Of course, for such an approach to be successful a
certain {\em locality} of the microphysics processes is necessary,
which can be expressed by the principle that neighboring material elements of a structure  ``communicate" only via forces at their common borders. 

A widely analogous procedure with respect to the task of calculating
transport coefficients for a certain type of turbulence would
consist in creating it in a (small) test volume with well defined
boundary/environmental conditions (like a penetrating magnetic and/or
gravitational field) and to determine then the wanted coefficients somehow
{\em by measurements}.
Then a major theoretical challenge consists in specifying the set of
experiments needed to find just these coefficients and in prescribing
the computational recipe for extracting them from the measured quantities
like fluctuating magnetic fields and/or velocities.

This program has been implemented by the so-called {\em test-field
methods} (Schrinner \ea 2005, 2007) with the modification that the
physical experiments are replaced by numerical ones and that occasionally,
in some of them, the ruling physics has been modified from the canonical one.
Clearly, the most important difference to the continuum mechanics scheme
lies in the fact that just the same equations whose direct simulation
was felt to be not affordable, and that just created the need for a MFM, have
now nevertheless to be simulated within the numerical experiments.
However, in two aspects the test-field approach can still be advantageous.
Firstly, the ``experimental" volumes can represent small sections
of the object which is to be globally analyzed.
Hence, much finer structures can be resolved for the same numerical
effort.
Secondly, if a MFM is established once, it can thereafter
be utilized for long-term global
simulations which would otherwise be prohibitively expensive.

The aforementioned locality of the actual physics  has here  to be required
with respect to correlation properties of the underlying fluctuating
fields, say a turbulent flow.
For all conceivable astrophysical situations this condition can hardly be
considered too restrictive.
In practice, correlation lengths and times are the relevant quantities
to be considered in defining the simulation box size and the integration
time.

Computing turbulent transport
tensors like $\aTens$ and $\eTens$ (see Eq.~(\ref{Eansatz}) below)
has now been done with variable success over the last 20 years.
Utilizing the imposed-field method for $\aTens$
has either led to the confirmation of well-known results (for
example a positive horizontal $\alpha$ effect in the upper layers of
convection in the Northern hemisphere), or to the prediction of as yet
unknown results (e.g.\ a reversed sign of the vertical $\alpha$ under
the same conditions; see Brandenburg \ea 1990) later
confirmed by theoretical calculations (Ferri\`ere 1992,
R\"udiger and Kitchatinov 1993).

After having explained the new test-field method in the next section,
particular applications considering the non-locality
of turbulent transport in space and time will be discussed in \Sec{Nonlocality}.

\subsection{Test-field method}
\label{TestField}
Let us return to \Eq{dbdt} for the fluctuating magnetic field.
The wanted mean electromotive force $\meanemf=\overline{\uu\times\bb}$
is obviously a linear and homogeneous functional of 
$\meanBB$ and we may therefore write the ansatz 
\begin{equation}
\meanemf=\aTens\meanBB - \eTens\nab \meanBB, \label{Eansatz} 
\end{equation}
strictly valid for stationary mean fields depending only linearly on position.
The components of $\aTens$ and $\eTens$ can be found by the following procedure:
\begin{enumerate}
\item 
solve 
\[
\parder{\bb^k}{t} -\eta\nabla^2\bb^k-\curl\left[
\meanUU\times\bb^k +(\uu\times\bb^k)'\right] = \curl(\uu\times\meanBB^k)
\]
for given $\uu,\meanUU$ and {\em test fields} $\meanBB^k, \; k=1,\ldots,N$,
\item
calculate $\meanemf^k=\overline{\uu\times\bb^k}$,
\item
determine the components of $\aTens$, $\eTens$ by inverting
\EQ
\meanemf^k=\aTens\meanBB^k - \eTens\nab \meanBB^k. \label{ansatz}
\EN
\end{enumerate}
The solution is unique, if
$N$ chosen appropriately  and the 
test fields $\meanBB^k$ are sufficiently independent.
Since we ``look at" the given velocity fields $\uu$, $\meanUU$ not only
from one perspective like in the case of the imposed-field method,
but obtain instead different views represented by the different test
solutions $\bb^k$, the test-field approach could be characterized as
``holographic" instead of ``photographic".
Indeed, the whole information needed to specify an ansatz like (\ref{ansatz})
is extracted from the velocity fields.

For a number of explicitly given flows like those introduced by
Roberts (1970) and Galloway and Proctor (1992),
exact agreement of the determined tensors with SOCA results.
In the case of the Roberts flow there is agreement even with an analytic
result for arbitrary magnetic Reynolds numbers (R\"adler \ea 2002,
Rheinhardt and Brandenburg 2010, R\"adler and Brandenburg 2009).

The method has now been applied to a number of different flows ranging
from homogeneous forced turbulence without shear (Sur \ea 2008,
Brandenburg \ea 2008a) to cases with shear (Brandenburg \ea 2008b,
Mitra \ea 2009) and to inhomogeneous turbulence in stratified disks
(Brandenburg 2005a, Gressel \ea 2008) as well as convection
(K\"apyl\"a \ea 2009a).

\subsection{Nonlocality in space and time}
\label{Nonlocality}

Because the test fields are not native to the system, they can disentangle effects
that can not otherwise be distinguished.  However,
for the same reason
they can introduce temporal or spatial scales that are
again not native to the system.

With respect to temporal scales the consequences of this mismatch
can be seen in the  {\em memory effect}:
Consider a dynamo system with a growing mean field.
Turbulence creates a fluctuating field from the mean one, and as long as
it
survives it contributes to the electromotive force.  If the mean
field is
growing,
the small scale field created in the past is weaker than it would
be
when
created in the present. Thus,  if the time behavior of the test fields is different
from that of the ``real'' mean field, the
$\aTens$ and $\eTens$ tensors
from the test-field method will not be
the actual ones which rule the evolution of the mean field.
A similar
problem occurs when the spatial scales of the test-fields do not coincide with
the spatial scale of the mean field to be modeled,
due to
{\em non-locality} in space.

When the proper scales of the system are known, corresponding
test-fields can be used.
Otherwise, the scales of the test fields, say, wavevector $\kk$ and
frequency $\omega$ are considered independent parameters and the
test-field method provides $\aTens(\kk,\omega)$ and $\eTens(\kk,\omega)$
which exhaustively describe the response kernel introduced in (\ref{convol}).

\begin{figure}
\centering
\includegraphics[width=\textwidth]{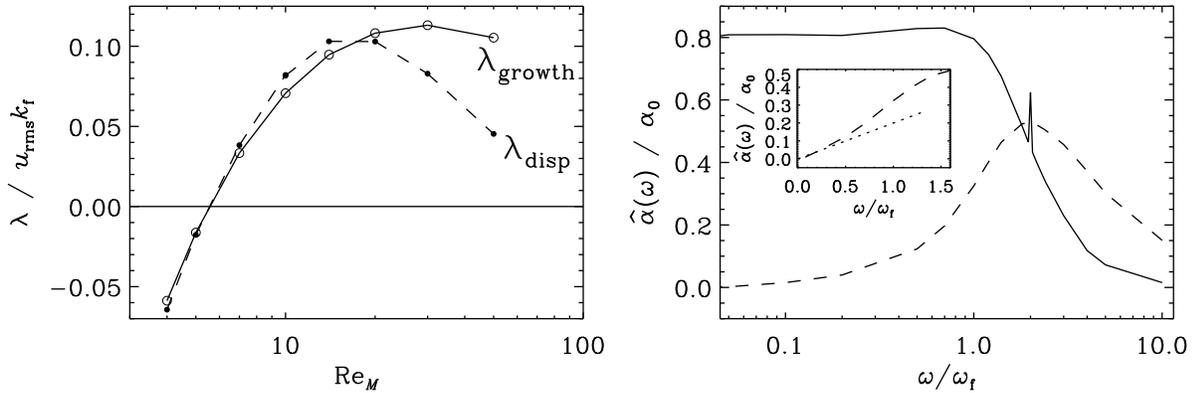}
\caption{
Left:
$\Rm$ dependence of the dynamo growth rate for the Roberts flow as obtained
from a direct calculation ($\lambda_{\rm growth}$) compared with the
result of the dispersion relation,
$\lambda_{\rm disp}=|\alpha k|-(\eta+\etat)k^2$,
using a cube of size $L^3$; $k_1=2\pi/L$, $\kf=\sqrt{2}k_1$.
For this range of $\Rm$, the most unstable mode
has the largest possible wavelength ($k=k_1$).
Right: 
Real and imaginary parts of $\tilde\alpha(\omega)$ for $k=0$
using the Otani (1993) MW+ flow with $\Rm=1$.
Normalization given by $\alpha_0=u_0$.
Inset: 
scaling of $\mbox{Im}\,\tilde\alpha$ near the origin
with slope 0.2, in agreement with the results of Hughes and Proctor (2010).
Adapted from Hubbard and Brandenburg (2009).
}
\label{Fig-memory}
\end{figure}

The memory effect is demonstrated
in \Fig{Fig-memory} for the case of the
Roberts flow (for details, see Hubbard and Brandenburg 2009).
In the left plot, we see the difference between the growth rate of a dynamo
to that calculated from the dispersion relation using
$\aTens$ and $\eTens$ determined by the test-field method
with steady test fields.
We can reconcile these growth rates 
by deriving them all from a proper kernel which can be established by employing
a set of test fields with different time dependencies.

The memory effect and non-locality in space have been studied using the integral kernel
technique in Hubbard and Brandenburg (2009) and Brandenburg \ea (2008a), respectively.
Using test-fields that oscillate
sinusoidally in time, the Fourier transforms
$\hat{\alpha}(\omega)$ of the turbulent transport integral kernels
$\tilde{\alpha}(t)$ were found to fit the form:
\EQ
\frac{\hat{\alpha}(\omega)}{\alpha_0}
=\frac{1-\ii \omega \tau_\alpha}{(1-\ii \omega \tau_\alpha)^2
+\omega_\alpha^2 \tau_\alpha^2},\label{kernel}
\EN
where $\tau_\alpha$ is the memory time of the flow and $\omega_\alpha$ 
is a fit parameter of order $\tau_\alpha^{-1}$.
In turbulence $\tau_\alpha$ is comparable to the turn-over time
but in steady flows it can approach microphysical resistive time scales. 
In the right panel of \Fig{Fig-memory} we present such a fit
for the MW+ flow of Otani (1993),
with an extra spike at $\omega/\omega_{\rm f}=2$, a natural frequency of
this flow.
The slope of the imaginary part at the origin,
$\dd\mbox{Im}\,\hat\alpha/\dd\omega|_{\omega=0}$, 
represents the coefficient of the first order term with respect to
an expansion in time.
Its value of 0.2 is in agreement with that found by Hughes and Proctor (2010).
The real-space integral kernel corresponding to (\ref{kernel}) reads
\EQ
\tilde{\alpha}(\tau)
=\alpha_0 \Theta(\tau) e^{-\tau/\tau_\alpha} \cos \omega_\alpha \tau,
\EN
where $\tau$ is the time distance to the instant of consideration, and
$\Theta$ is the Heaviside step function that preserves causality
as the time integral kernel must only consider the past.

\begin{figure}
\centering
\includegraphics[width=\textwidth]{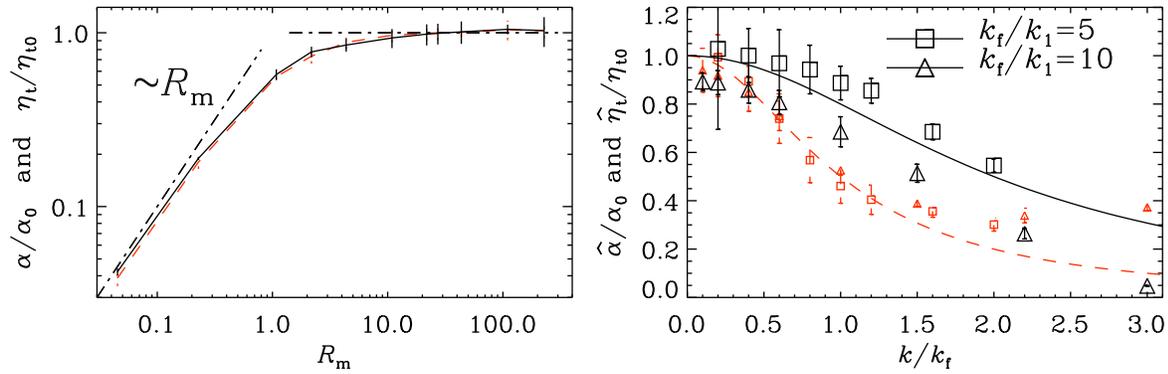}
\caption{
Left:
Dependence of the normalized values of $\alpha$ (dashed or red line)
and $\eta_{\rm t}$ (solid line)
on $\Rm$ for $k/\kf=0.2$ and $\mbox{Re}=2.2$.
Adapted from Sur \ea (2008).
Right:
Dependences of the normalized $\tilde\alpha$
(dashed or red line, small symbols)
and $\tilde\eta_{\rm t}$ (solid line, bigger symbols)
on the normalized wavenumber $k/k_{\rm f}$
for turbulence forced
with  $k_{\rm f}/k_1=5$, $\Rm=10$ (squares)
and $k_{\rm f}/k_1=10$, $\Rm=3.5$ (triangles).
Lines give the Lorentzian fits \eq{KernelsTurb}.
Adapted from Brandenburg \ea (2008a).
}
\label{Fig-spatial}
\end{figure}

The Fourier transform of the spatial integral kernels
is somewhat simpler, being fit by a Lorentzian:
\EQ
\frac{\hat{\alpha}(k)}{\alpha_0}=\frac{1}{1+(a_\alpha k/\kf)^2}
\quad\mbox{and}
\quad \frac{\hat{\eta}_t(k)}
{\eta_{t0}}=\frac{1}{1+(a_\eta k/\kf)^2},
\label{KernelsTurb}
\EN
whose amplitude is nearly independent of $\Rm$ for $\Rm\gg1$
and the $k$ dependence is reasonably close to quadratic for $k/\kf<2$;
see \Fig{Fig-spatial}.
Here, $a_\alpha\approx2a_\eta\approx1$ are coefficients of order unity.
The corresponding spatial integral kernels are simple exponential decays:
\EQ
\tilde{\alpha}(\zeta)={\alpha_0\kf\over2a_\alpha}
e^{-(\kf/a_\alpha)\zeta}\quad\mbox{and}
\quad \tilde{\eta}_t(\zeta)={\eta_{t0}\kf\over2a_\eta}
e^{-(\kf/a_\eta)\zeta},
\EN
where $\zeta$ is the measure of distance from the point of consideration.

\section{From linear to nonlinear}
\label{nonlin}
As outlined in \Sec{TestField}, a mathematically well-founded procedure is available 
by which the tensors $\aTens$ and $\eTens$ can be obtained when the velocity 
is given.
Naturally, the question arises of how one should proceed when the
mean magnetic field has already acted upon this velocity.
 Inspecting \Eq{dbdt}
it can be concluded that $\meanBB$,
considered as a functional of $\UU$ and $\meanBB$,
is always linear and homogeneous in the latter, irrespective of the effects $\UU$ was 
subjected to, and whether a mean field had already acted upon it or not.
That is, the presented method continues to be valid without  modification and as a tribute to this
extension it is called the {\em quasi-kinematic method}.
The turbulent transport coefficients are of course now depending on $\meanBB$, but this dependence
is entirely conveyed by the dependence of $\UU$ on $\meanBB$.

\begin{figure}
\centering
\includegraphics[width=\textwidth]{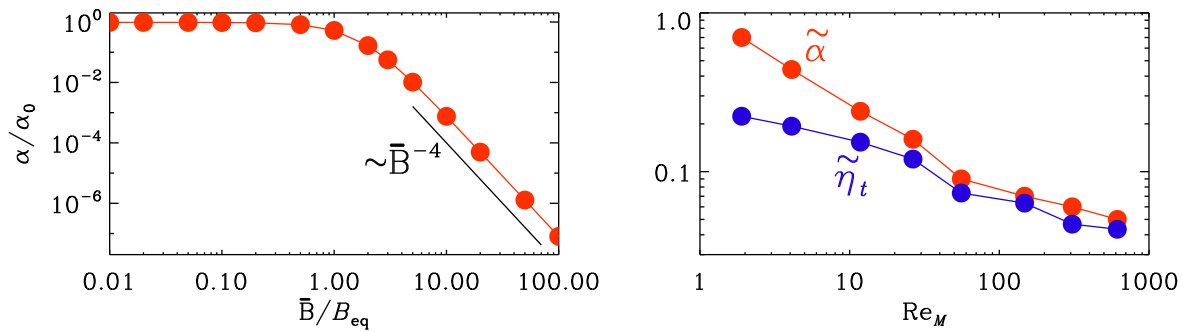}
\caption{\label{quasikin}
Left:
Variation of $\alpha$ with $\meanB$ for the driven Roberts flow
with $\Rm=1/2\sqrt{2}\approx0.35$ and $\Pm=1$.
Adapted from Rheinhardt and Brandenburg (2010).
Right: 
$\Rm$-dependence of the normalized $\tilde\alpha$ and $\tilde\etat$
in the saturated state with $\meanB\approx\Beq$.
Adapted from Brandenburg \ea (2008c).
}
\end{figure}

Clearly, a dynamically effective mean field now represents an additional
preferred direction.
As a consequence, for an isotropic hydrodynamic background and a uniform
$\meanBB$, the formerly isotropic tensor $\aTens$ adopts now the shape
\EQ
\alpha_{ij} = \alpha_1 \delta_{ij}  + \alpha_2 \hatB_i \hatB_j,
\quad i,j=1,2, \label{nonlinalp}
\EN
with $\hatBB$ the unit vector in the direction of $\meanBB$.
If this is, say, the $x$ direction we get 
$\alpha_{11}=\alpha_1+\alpha_2$ and $\alpha_{22}=\alpha_1$.
Both coefficients are of course functions of $\meanB$.
Since $\meanEE=\alpha_{11} \meanBB$, the effective
scalar $\alpha$ effect is just given by $\alpha_{11}$. 
As an example, the $\alpha$ quenching characteristic for the Roberts flow
was determined exhibiting a $\meanB^{-4}$ asymptotic dependency,
cf.\ Rheinhardt and Brandenburg (2010) and \Fig{quasikin}.
This result is at odds with theoretical predictions, although it agrees
with data for a forced ABC flow (Sur \ea 2007).

Things become more involved if the direction of
the mean current density $\meanJJ$
enters as a second preferred direction.
A situation in which this complication is circumvented, without being
as simple as the former one, is given by the $\alpha^2$ dynamo due to
homogeneous isotropic helical (forced) turbulence.
Here, the dynamo solution is a Beltrami field with
$\meanBB\parallel\meanJJ$ and constant modulus.
Hence, $\meanJJ$ is not providing an additional preferred direction
and the form of \Eq{nonlinalp} remains valid.
The solution has always Beltrami shape, regardless at what level it
eventually saturates.
Consequently, $\alpha$ and $\etat$ are independent of position for any
field strength.
The growth rate of the dynamo is given by 
$\lambda= |\alpha k| - (\eta+\etat)k^2$
and should approach zero in the course of saturation.
Under these conditions it is possible to confirm the quasi-kinematic
method in the way that  $\alpha(\meanB)$ and $\etat(\meanB)$ are
determined in the saturated stage and checked for consistency against
$\lambda=0$.
Indeed, this could be demonstrated to high accuracy for different
values of $\Rm$; see Brandenburg \ea (2008c).
\FFig{quasikin} shows $\alpha(\meanB)$ and $\etat(\meanB)$
as functions of $\Rm$ in the saturated state with $\meanB\approx\Beq$.

\section{Quasi-kinematic method for magnetic buoyancy-driven flows}
\label{quasi}

So far we have been dealing with situations in which a hydrodynamic background
was provided independently and the mean magnetic field occurred as an
additional, at most coequal participant.
But what about cases in which the turbulence itself is a consequence of $\meanBB$?
Examples are the magneto-rotational instability and the
magnetic buoyancy instability (see below).
Clearly, those setups do not know a kinematic stage
on which the influence of $\meanBB$ is negligible.
One might worry that in such a situation the quasi-kinematic
test-field procedure fails (Courvoisier \ea 2010).
However,
\Eq{dbdt} continues to be valid and hence all conclusions drawn from it.
Consequently, the quasi-kinematic method should be applicable.
The only peculiarity occurring here is the fact that
all components of $\aTens$ and $\eTens$
vanish for $0\le \meanB \le \meanB_{\mathrm{threshold}}$, because a
fluctuating velocity (and magnetic field) develops only after the
instabilities have set in.

Let us now consider the magnetic buoyancy instability.
It has been proposed by Moffatt (1978) that,
once the dynamo-generated magnetic field 
in the overshoot layer of the Sun reaches appreciable strengths,
this instability can set in and govern the dynamics thereafter. 
The buoyancy instability of a
localized flux layer in the presence of stratification
and rotation was later studied in detail by Schmitt (1984, 1985).
A necessary but not sufficient condition for this instability is
\begin{equation}
\frac{\partial}{\partial z}\log\left(\frac{B^2}{\rho^2}\right) < 0,
\end{equation}
which essentially means that the magnetic field decreases faster with height than density. 
Using the imposed-field method,
Brandenburg and Schmitt (1998) performed numerical calculations
to determine the $\alpha$ effect of the resulting turbulence.
Here we determine all
components of $\aTens$ and $\eTens$ using the quasi-kinematic test-field method
wherein mean fields are defined as $xy$ averages.

Our setup is similar to that described in Brandenburg and Schmitt (1998).
The computational domain is a cuboid of size
$-1\le x\le 1$, $-3\le y\le 3$, $-0.5\le z\le 1.5$, with
gravity pointing in the negative $z$ direction,
and rotation $\OO$ making an angle $\theta$ with the vertical.
The pressure scale height is $H_{\rm P}=1$ (half height of the box). 
The base state is a polytrope with index $m=3$
(the adiabatic value is here 3/2), so that it is stable to convection.
The initial condition comprises
a horizontal magnetic layer of thickness $H_{\rm B}=0.1$ with the profile,
\begin{equation}
B_y= v_{\rm A0}H_{\rm B}\parder{}{z}\tanh\left(\frac{z-0.1}{H_{\rm B}}\right),
\end{equation} 
where the ratio of the Alfv\'en speed to the sound speed is
$v_{\rm A0}/c_{\rm s0}=0.5$. 
We modify the base state such that the density profile remains polytropic
but the entropy profile is adjusted to obey magnetostatic equilibrium.
The initial velocity
consists of about 20 localized eddies with Mach numbers of about $10^{-5}$ at 
$z=0.1$ in the $xy$ plane.
We use stress-free
boundary conditions for the velocity
and the vertical field condition for the magnetic
field, whereas with respect to entropy we keep
the temperature at the top and the (radiative) heat
flux at the bottom constant. 
All calculations have been done with $\Pr=\Pm=4$ on a $64^3$ grid.
Fig.~\ref{fig:bxurms}a gives the time series of the volume averages
$\bra{u^2}$ and $\bra{B_x^2}$ and Fig.~\ref{fig:bxurms}b
the evolution of the mean field $\overline{B}_x$.
There is a short exponential growth phase followed by
decay on a resistive time scale.

\begin{figure}
\label{fig:bxurms}
\includegraphics[width=0.333\textwidth]{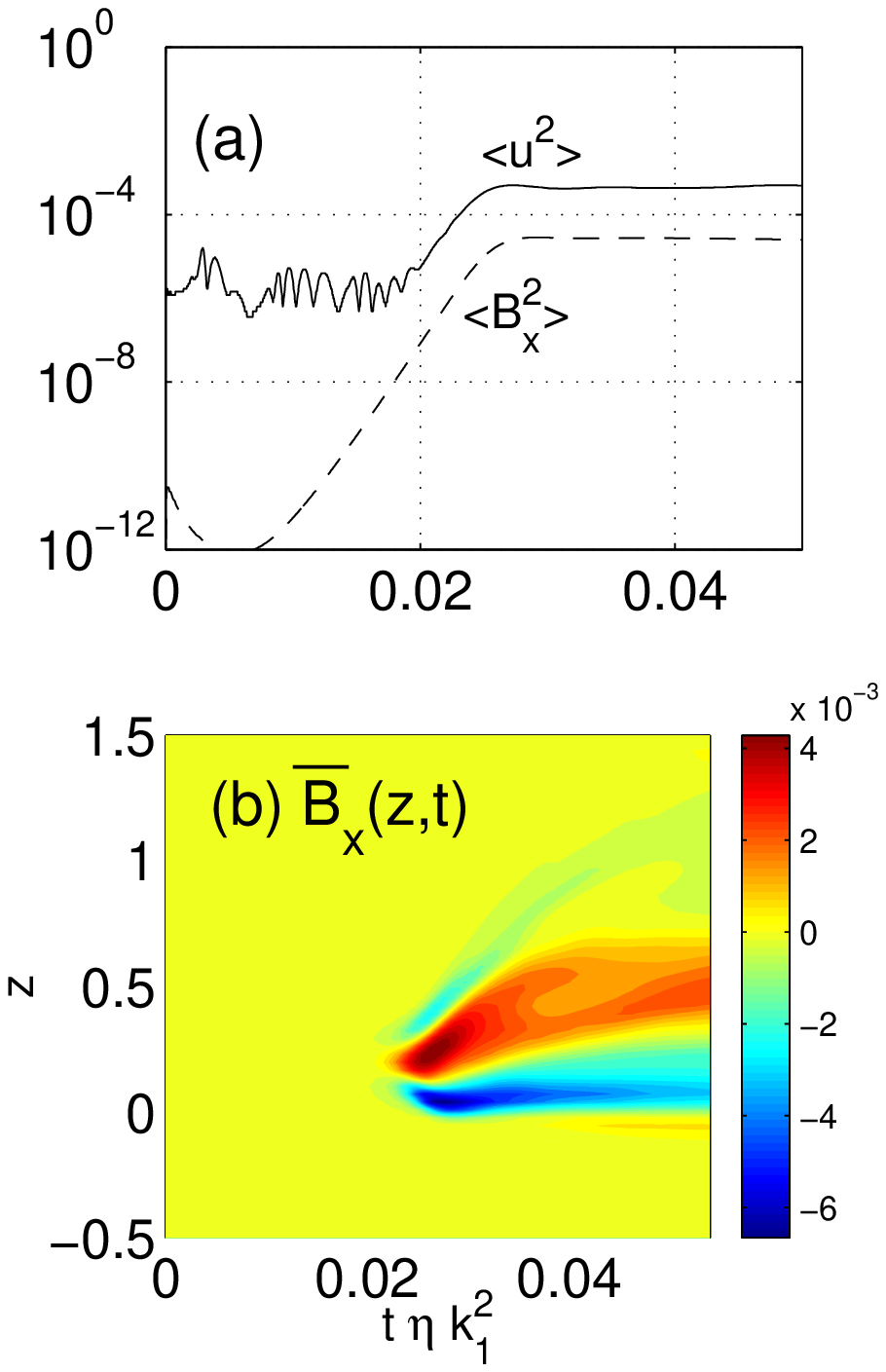}
\includegraphics[width=0.666\textwidth]{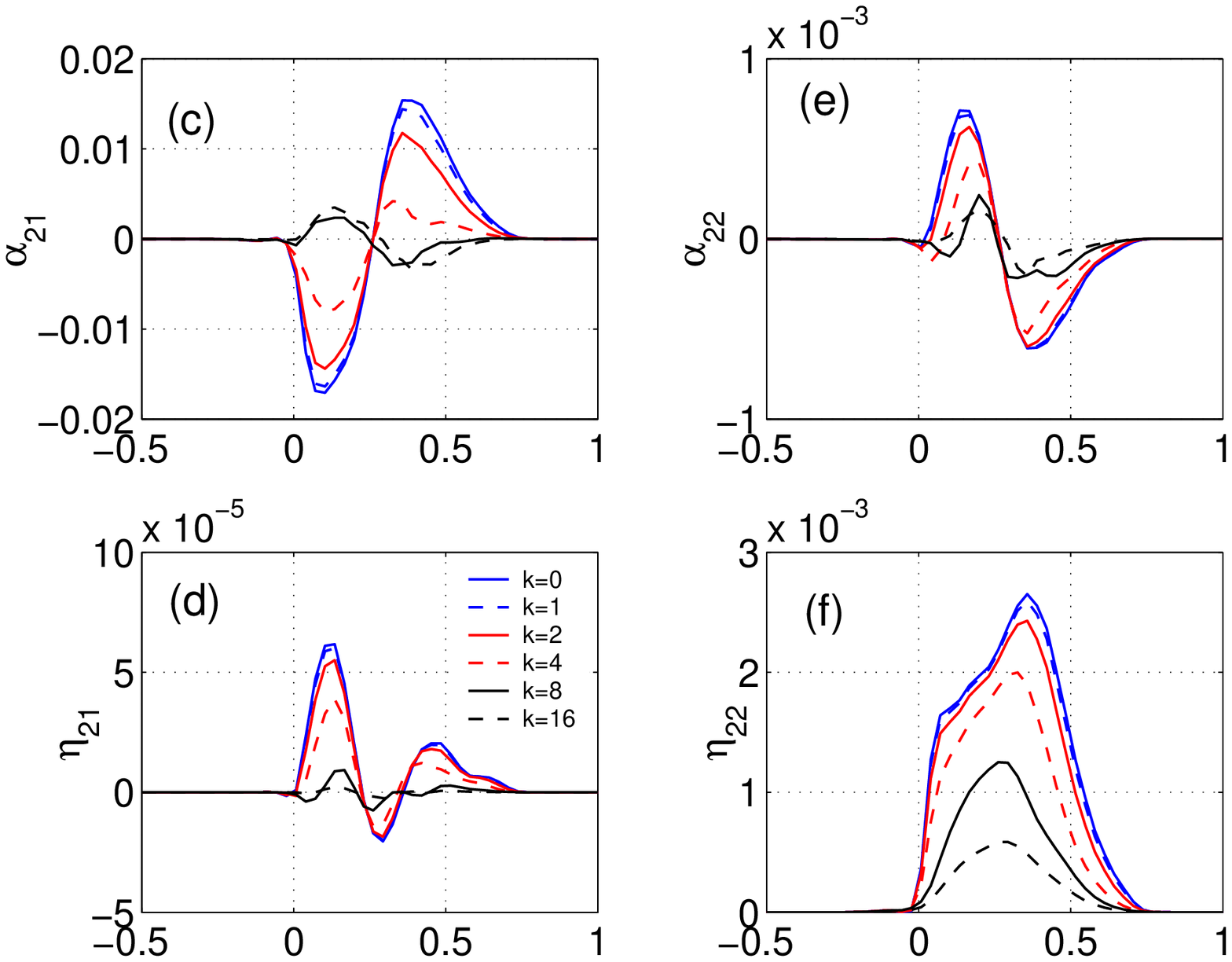}
{\caption{\label{fig:bxurms}
(a) Evolution of volume averaged
velocity and $B_x$ squares.
(b) A `butterfly diagram' for $\meanB_x$.
Panels (c) through (f) show selected tensorial components 
of $\aTens$ and $\eTens$ for different wavenumbers $k$
as explained in the legend in (d).
Plots have been obtained by time averaging over the interval indicated 
by the dashed-dotted lines in (a).}}
\end{figure}

When it comes to applying the test-field method, an aspect not discussed
up to now is the intrinsic 
inhomogeneity of the flow
both due to stratification and the background magnetic field itself.
Within kinematics, that is without the background field,
no specific complication is connected to this as
from the stationary version of \Eq{convol}
$\aTens$ and $\eTens$ emerge straightforwardly
in a shape expressing inhomogeneity, that is, 
$\aTens(\xx,\xx')$, $\eTens(\xx,\xx')$
or, equivalently, $\aTens(\xx,\xx-\xx')$, $\eTens(\xx,\xx-\xx')$.
When subjecting the latter to a Fourier transform with respect to their second argument,
we arrive at $\hat{\aTens}(\xx,\kk)$ and $\hat{\eTens}(\xx,\kk)$.
In our case, harmonic test fields with different wavenumbers $k$
in the $z$ direction can be employed to obtain $\hat{\aTens}(z,k)$
and $\hat{\eTens}(z,k)$.

In the nonlinear situation,
the Green's function approach remains valid if $\meanemf$ is considered
as a functional of $\UU$ and $\meanBB$ which is then linear and homogeneous in the latter (cf.\ \Sec{nonlin}).
However, we have to label $\boldsymbol{G}$ by the $\meanBB$
actually acting upon $\UU$,
that is, $\boldsymbol{G}(\xx,\xx';\meanBB)$, and can thus only make statements
about the transport tensors for just the particular $\meanBB$ at hand.
Hence, the tensors have to be labelled likewise: $\hat{\aTens}(z,k; \meanBB)$,
$\hat{\eTens}(z,k; \meanBB)$.
As our initial mean magnetic field is in the $y$ direction,
the instability will generate a $\meanB_x$
and we are mainly interested in
the coefficients
$\alpha_{yx}$, $\alpha_{yy}$, $\eta_{yx}$ and $\eta_{yy}$
with rank-2 tensor components $\eta_{ij}=-\eta_{ik3}\epsilon_{jk3}$;
they are shown in \Fig{fig:bxurms}, (c) through (f).

Our goal is now to confirm that the relationship between
$\meanemf$ and $\meanBB$ taken directly from the DNS
can be represented by \Eq{ansatz} with the transport tensors
determined using the quasi-kinematic test-field method.
In mathematical terms\\
\EQ
\meanemfs_i(z; \meanBB)\stackrel{\mathrm{\tiny ?}}{=}
\mathrm{Re}\Big\{\sum_k \left[\hat{\alpha}_{ij}(z,k; \meanBB)
-\ii k\hat{\eta}_{ij3}(z,k;\meanBB)\right] c_j^{(k)} \exp(\ii \pi k z)\Big\}
\label{assembly}
\EN
with $2\cc^{(k)} = \int \meanBB(z) \exp(-\ii \pi k z) dz$ and
$\stackrel{\mathrm{\tiny ?}}{=}$ signifying the question
whether both sides are indeed equal.
We find that we can reasonably reconstruct the mean emf by truncating the
infinite Fourier series already at $k=8$.
The result of the assembly of the $\meanEMF$
as formulated on the right hand side of (\ref{assembly}) is presented in 
Fig.~\ref{fig:emfconst}c and turns out to be a faithful reproduction
of $\meanemf$ shown in Fig.~\ref{fig:emfconst}a
especially during the exponential growth phase.
We conclude that the quasi-kinematic test-field method may be used
for correctly calculating transport coefficients even in the presence
of inhomogeneous turbulence driven by an initial mean magnetic field.

\begin{figure}
\label{fig:emfconst}
\includegraphics[width=0.9\textwidth]{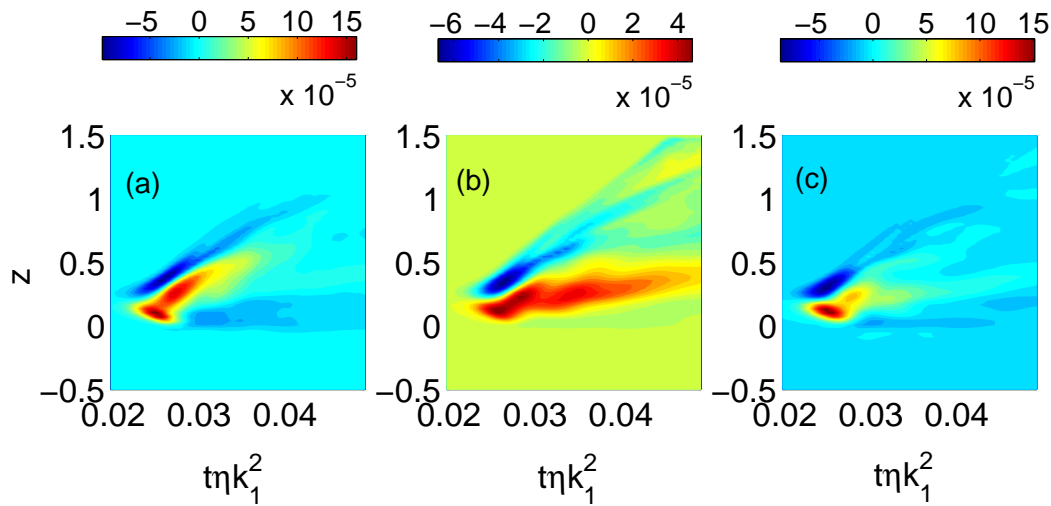}
\caption{\label{fig:emfconst}(a) The mean emf $\meanemfs_x(t,z)$
calculated from the horizontal average
$\overline{{\bm u}\times {\bm b}}$.
(b) Reconstruction of $\meanemfs_x(t,z)$
using only the $k=0$ contributions in (\ref{assembly}).
(c) Same as (b) but using all contributions $k=0, 1, ..., 8$.}
\end{figure}

\section{Conclusions}
Mean-field theory has still a lot to offer in terms
of new effects and quantitative precision by combining analytics with
numerics in parameter regimes that were previously inaccessible.
The list of examples goes on and on; here we have only mentioned
some of the most striking cases.
The unmentioned ones concern, e.g., the Reynolds and
Maxwell stresses that have important applications in accretion disks
(Blackman 2010) and possibly sunspot formation (Brandenburg \ea 2010).
One may anticipate that all these aspects
of mean-field theory will soon gain in
significance in our voyage toward understanding astrophysical dynamos.

\section*{Acknowledgements}

We acknowledge the allocation of computing resources provided by the
Swedish National Allocations Committee at the Center for
Parallel Computers at the Royal Institute of Technology in
Stockholm and the National Supercomputer Centers in Link\"oping
as well as the Norwegian National Allocations Committee at the
Bergen Center for Computational Science and the computing facilities
hosted by CSC - IT Center for Science Ltd.\ in Espoo, Finland,
who are administered by the Finnish Ministry of Education.
This work was supported in part by
the European Research Council under the AstroDyn Research Project
227952, the Swedish Research Council grant 621-2007-4064, and the
Finnish Academy grant 121431.

\section*{References}

\bibliographystyle{aa}

\end{document}